\documentclass[12 pt,twoside,aps,prd,amsmath,amssymb,
tightenlines,showpacs,showkeys,eqsecnum]{revtex4}

\usepackage{graphicx}
\usepackage{mathptmx}
\usepackage{bm}
\usepackage{amssymb}
\usepackage{amsmath}
\def\myfigure #1#2#3#4
{\begin{figure}[ht]\begin{center}
\includegraphics[width=#2 \textwidth]{#1.jpg}
\parbox[t]{#4\textwidth}{\caption{#3}\label{#1}}
\end{center}\end{figure}}

\def \myfigures #1#2#3#4#5#6#7#8
{\begin{figure}[ht]
    \begin{center}
        \includegraphics[width=#2 \textwidth]{#1.jpg}
        \hfill
        \includegraphics[width=#6 \textwidth]{#5.jpg}
        \parbox[t]{#4\textwidth}{\caption {#3}\label{#1}}
        \hfill
        \parbox[t]{#8\textwidth}{\caption {#7}\label{#5}}
    \end{center}
\end{figure} }

\begin{document}
\title{Static spherically symmetric
space-time: some remarks}
\author{Bijan Saha}
\affiliation{Laboratory of Information Technologies\\
Joint Institute for Nuclear Research, Dubna\\
141980 Dubna, Moscow region, Russia\\ and\\
Institute of Physical Research and Technologies\\
RUDN University\\
Moscow, Russia} \email{bijan@jinr.ru}
\homepage{http://spinor.bijansaha.ru}

\hskip 1 cm

\begin{abstract}
Within the scope of a spherically symmetric space-time we study the
role of different types of matter in the formation of different
configurations with spherical symmetries. Here we have considered
matter with barotropic equation of state, scalar field,
electromagnetic field and an interacting system of scalar and
electromagnetic field as the source. Corresponding field equations
are solved exploiting harmonic coordinates. An easy to handle method
is proposed which allows one to have an idea about the possible
behavior of the metric functions once the components of the EMT of
the source field is known.
\end{abstract}



\maketitle

\section{Introduction}

In order to describe simple isolated bodies and island-like
configurations spherical symmetry is a natural choice
\cite{BronBook}. Spherically symmetric space-times are invariant
under spatial rotation. Metric functions in this case, generally,
depend on the radial coordinate and the time coordinate. In case of
a static space-time, metric functions do not depend on time.

Static spherically symmetric space-time is widely used in physics to
obtain analytic and numerical solutions to the Einstein field
equations in presence of different types of source fields. One of
the most celebrated static spherically symmetric solutions to the
Einstein equations is the Schwarzschild solution. Since the metric
functions depend on only the radial coordinate the static
spherically symmetric space-time gives rise to a simpler system of
equations which is easier to analyze. This type of space-time is
used to study different types of astrophysical objects such as black
holes, wormholes, compact stars etc. Since there is a very large
number of studies in this area, we mention just a few. Role of the
spinor field in the formation of black hole and/or wormhole was
investigated in \cite{Saha2018,BRS2020}. The stability of static,
spherically symmetric solutions of Rastall's theory was studied in
\cite{Bron2020}.

The aim of this paper is to analyze the static spherically symmetric
space-time in the presence of different matters with typical
energy-momentum tensor (EMT) widely exploited in literature. In our
view, the method, offered here, will be useful to solve the system
of equations in question both analytically and numerically. It will
allow the researchers to impose minimal additional conditions to
obtain the solutions.

\section{Basic Equation}

The action we choose in the form

\begin{eqnarray}
{\cal S} = \int \sqrt{-g} \left[\frac{R}{2 \kappa} + L \right] d
\Omega. \label{action}
\end{eqnarray}

where $\kappa = 8 \pi G$  is Einstein's gravitational constant, $R$
is the scalar curvature and $L$ is the matter field Lagrangian. We
don't specify the Lagrangian right now. It might be given by the
spinor, scalar, electromagnetic field or their interaction. We
specify it in the course of our journey.

The spherically symmetric metric we choose in the form
\begin{equation}
ds^2 = e^{2 \gamma} dt^2 - e^{2 \alpha} du^2 - e^{2 \beta}
(d\vartheta^2 + \sin^2{\vartheta} d \varphi^2), \label{ss}
\end{equation}
where the metric functions $\gamma, \alpha, \beta$ depend on the
spatial variable $u$ only. Since in order to describe the
spherically symmetric gravitational field we need only two
components of the metric tensor \cite{shikin}, then in \eqref{ss} it
is possible to choose explicitly one of the three metric functions
$\gamma, \alpha, \beta$ or demand that all these functions satisfy
one of the following coordinate conditions \cite{BronBook,shikin}:
\begin{enumerate}
\item $\alpha = 0$, i.e. $e^\alpha = 1$ - the Gaussian normal coordinates;
\item $\alpha = \gamma$ - isometric or tortoise coordinates;
\item $\alpha = \beta$ - homogeneous coordinates;
\item $e^{2\beta} = e^{2\alpha} u^2$ - isotropic coordinates;
\item $e^{\beta} = r$ - curvature or Schwarzschild coordinates. $r$ is the radius of
the sphere with $u$ = const. In this case the metric \eqref{ss}
takes the form

\begin{equation}
ds^2 = e^{2 \gamma (r)} dt^2 - e^{2 \alpha (r)} dr^2 - r^2
(d\vartheta^2 + \sin^2{\vartheta} d \varphi^2). \label{ss1}
\end{equation}

\item $\alpha = \gamma + 2 \beta $ - harmonic coordinates;
\item $\alpha = - \gamma$ - quasiblobal coordinates.
\end{enumerate}

In should be noted that since we consider the static spherically
symmetric configuration, all the field functions should depend on
the spatial variable $u$ only.

The Einstein tensor corresponding to the metric \eqref{ss} possesses
only diagonal elements, hence the Einstein equations in this case
takes the form

\begin{subequations}
\label{EET}
\begin{eqnarray}
\left( 2 \gamma^\prime \beta^\prime +
\beta^{\prime 2}\right) - e^{2 (\alpha - \beta)} &=& - \kappa T_1^1, \label{EE11}\\
\left(\gamma^{\prime 2} + \gamma^\prime \beta^\prime -\gamma^\prime
\alpha^\prime + \beta^{\prime 2} -\beta^\prime \alpha^\prime +
\gamma^{\prime\prime} +
\beta^{\prime\prime}\right) &=& -\kappa T_2^2, \label{EE22}\\
\left(3 \beta^{\prime 2} - 2 \beta^\prime \alpha^\prime +
2\beta^{\prime\prime}\right)- e^{2 (\alpha - \beta)} &=& - \kappa
T_0^0. \label{EE00}
\end{eqnarray}
\end{subequations}

Subtraction of \eqref{EE11} from \eqref{EE00} gives

\begin{equation}
\beta^{\prime \prime} + \beta^{\prime 2} - \alpha^\prime
\beta^\prime - \gamma^\prime \beta^\prime = - \frac{\kappa}{2}
\left[T_0^0 - T_1^1\right]. \label{0-1}
\end{equation}
subtraction of \eqref{0-1} from \eqref{EE22} yields

\begin{equation}
\gamma^{\prime \prime} + \gamma^{\prime 2} - \alpha^\prime
\gamma^\prime + 2 \gamma^\prime \beta^\prime =  - \frac{\kappa}{2}
\left[2 T_2^2 - T_0^0 + T_1^1\right]. \label{2-0-1}
\end{equation}

For numerical study it is convenient to rewrite the equations
\eqref{0-1} and \eqref{2-0-1} in the Cauchy form:

\begin{subequations}
\label{CF}
\begin{align}
\beta^\prime &= \nu, \label{CF1}\\
\gamma^\prime &= \tau, \label{CF2}\\
\nu^\prime + \nu^2 - \alpha^\prime \nu - \nu \tau &= -
\frac{\kappa}{2} \left[T_0^0 - T_1^1\right], \label{CF3}\\
\tau^\prime + \tau^2 - \alpha^\prime \tau + 2 \nu \tau &= -
\frac{\kappa}{2} \left[2 T_2^2 - T_0^0 + T_1^1\right]. \label{CF4}
\end{align}
\end{subequations}

To solve this system we have to know the concrete form of
energy-momentum tensor and some additional relation which is known
as coordinate condition. One can exploit one of the coordinate
conditions listed above. Note that in many problems considered
within the scope of static spherically symmetric space-time we deal
with the EMT such that

\begin{eqnarray}
T_0^0 &=& T_2^2 = T_3^3, \nonumber\\
T_1^1 &=& T_2^2 = T_3^3, \nonumber\\
T_0^0 &=& T_1^1 = - T_2^2 = -T_3^3. \nonumber
\end{eqnarray}

So in any of those cases listed above the task becomes even easier.

In what follows we consider the harmonic radial coordinate such that
$\alpha = \gamma + 2 \beta$. In view of it the Eqns. \eqref{CF3} and
\eqref{CF4} can be written as

\begin{subequations}
\label{CFhc}
\begin{align}
\nu^\prime &=  \nu^2 + 2 \nu \tau -
\frac{\kappa}{2} \left[T_0^0 - T_1^1\right], \label{CF3hc}\\
\tau^\prime &= - \frac{\kappa}{2} \left[2 T_2^2 - T_0^0 +
T_1^1\right]. \label{CF4hc}
\end{align}
\end{subequations}

Now, to find the metric functions we have to know the
energy-momentum tensor (EMT) $T_\mu^\nu$ of the material field.
Depending on the source fields it may vary. Further we consider a
few cases those are widely used in literature both listed above and
beyond.

\vskip 5mm

{\bf Perfect fluid}

\vskip 3mm

The first case we consider is the one when the source field is given
by $T_\nu^\mu = {\rm diag}{(\varepsilon,\, -p,\,-p,\,-p)}$, i.e.
$T_1^1 = T_2^2 = T_3^3$ . In this case from the conservation law
\begin{equation}
T_{\nu;\mu}^{\mu} = T_{\nu,\mu}^{\mu} + \Gamma_{\rho \mu}^{\mu}
T_{\nu}^{\rho} - \Gamma_{\nu \mu}^{\rho} T_{\rho}^{\mu} = 0,
\label{cl}
\end{equation}
in view of the fact that $\varepsilon$ and $p$ depends on $u$ we
find
\begin{equation}
p^\prime + (p + \varepsilon) \gamma^\prime = 0. \label{cl2}
\end{equation}
The perfect fluid obeys the barotropic equation of state (EOS) such
that
\begin{equation}
p = W \varepsilon, \label{bes}
\end{equation}
where $W$ is a constant and known as EOS parameter. For $W \ge 0$ it
describes a perfect fluid such as dust, radiation, hard Universe and
stiff matter. In case of $W < 0$ we have dark energy such as
quintessence, $\Lambda$ term and phantom matter.

In this case from the \eqref{cl2} one finds
\begin{equation}
\varepsilon = \left(C e^{-\gamma}\right)^{(1+1/W)}.
\label{varepsilon}
\end{equation}
Here $C$ is the integration constant and can be taken to be unity.
In this case we have the following system of equations

\begin{subequations}
\label{CFhcpf}
\begin{align}
\beta^\prime &= \nu, \label{CFpf1}\\
\gamma^\prime &= \tau, \label{CFpf2}\\
\nu^\prime &=  \nu^2 + 2 \nu \tau - \frac{\kappa}{2}(1-W) e^{-(1 +
1/W)\gamma}, \label{CFpf3hc}\\
\tau^\prime &= -(3W - 1) e^{-(1 + 1/W)\gamma}. \label{CFpf4hc}
\end{align}
\end{subequations}

We solve the system \eqref{CFhcpf} numerically. For simplicity we
set $\kappa = 1$. The initial values are taken to be $\tau (0) =
0.2,\, \nu (0) = 0.2,\, \gamma (0) = 0.3,\, \beta (0) = 0.3$. In
Figs. \ref{metricpfrad} and \ref{metricpfquint} we have plotted the
behavior of metric functions for $W = 1/3$ which corresponds to
radiation and $W=-2/3$ that corresponds to quintessence,
respectively. In the first case $\beta (u)$ is growing faster than
$\gamma (u) $, whereas in the second case $\beta (u)$ is decreasing
as $u$ increase while $\gamma (u)$ is still increasing.

\vskip 1 cm \myfigures{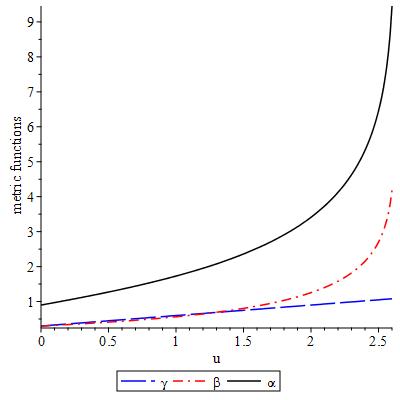}{0.46}{Plot of metric functions
for $W = 1/3$}{0.45}{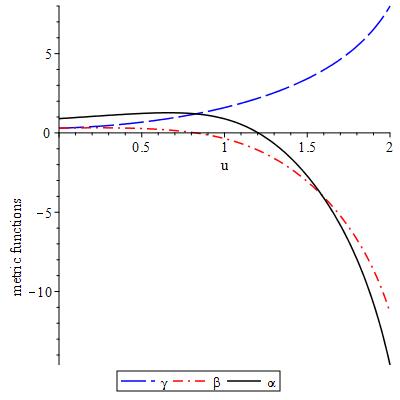}{0.43}{Plot of metric functions
for $W = -2/3$}{0.45}

\vskip 5mm

{\bf Scalar field }

\vskip 3mm

Let us consider a scalar field with the Lagrangian

\begin{equation}
L_{\rm sc} = \frac{1}{2} \varphi_{,\alpha} \varphi^{,\alpha} -
V(\varphi). \label{sc}
\end{equation}
If $V(\varphi) = (1/2) m^2 \varphi^2$ the foregoing Lagrangian leads
to the Klein-Gordon equation. The corresponding EMT reads

\begin{equation}
T_{{\rm sc} \nu}^{\mu} = \varphi_{,\nu} \varphi^{,\mu} -
\delta_\nu^\mu (\frac{1}{2} \varphi_{,\alpha} \varphi^{,\alpha} -
V(\varphi)). \label{emtsc}
\end{equation}
If the scalar field depends on only redial coordinate $u$ we find
$T_0^0 = T_2^2 = T_3^3 =  e^{-2\alpha} \varphi^{\prime 2} +
V(\varphi)$ and $T_1^1 = - e^{-2\alpha} \varphi^{\prime 2} +
V(\varphi).$

The corresponding scalar field equation
\begin{equation}
\partial_\mu\left(\sqrt{-g} g^{\mu\nu} \varphi_{,\nu}\right) +
\sqrt{-g} V_\varphi = 0, \quad V_\varphi = \partial V/\partial
\varphi. \label{scfe}
\end{equation}

On account of coordinate condition in this case we finally have the
following system of gravitational and scalar field equations:

\begin{subequations}
\label{CFhcsc}
\begin{align}
\beta^\prime &= \nu, \label{CFsc1}\\
\gamma^\prime &= \tau, \label{CFsc2}\\
\nu^\prime &=  \nu^2 + 2 \nu \tau + \kappa e^{-2 (\gamma + 2 \beta)}
\psi, \label{CFsc3hc}\\
\tau^\prime &= - \kappa V \label{CFsc4hc}\\
\varphi^\prime &= \psi, \label{psi}\\
\psi^\prime &= e^{\gamma + 2 \beta} V_\varphi. \label{sc}
\end{align}
\end{subequations}

The system \eqref{CFhcsc} we solve numerically. As in previous case
we set $\kappa = 1$. The initial values are set to be $\tau (0) =
0.05,\, \nu (0) = 0.05,\,\gamma (0) = 0.5,\, \beta (0) = 0.5,\,
\varphi (0) = 0.1,\, \varphi^\prime (0) = 0.2$. In Fig.
\ref{metricscalar} we have drawn the picture of the metric functions
when the source is given by the scalar field. In Fig. \ref{scalar}
the corresponding scalar field is demonstrated. As one sees, in this
case $\gamma (u)$ decreases with the growth of $u$, while $\beta
(u)$ slowly increases.

\vskip 1 cm \myfigures{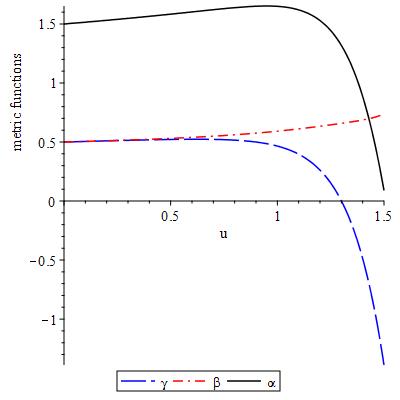}{0.46}{Plot of metric functions
for $V(\varphi) = 2 m^2 \varphi^2$ }{0.45}{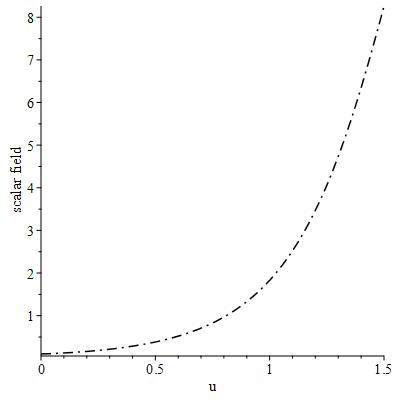}{0.43}{Plot of
scalar field satisfying equation \eqref{scfe}}{0.45}

Note that in case of spinor field we also obtain $T_0^0 = T_2^2 =
T_3^3$  \cite{Saha2018,BRS2020} and the corresponding solutions will
be similar to the one found above.

\vskip 5mm

{\bf Electromagnetic field }

\vskip 3mm

Let us now consider the case with electromagnetic field with the
Lagrangian

\begin{equation}
L_{\rm em} = \frac{1}{4} F_{\mu\nu} F^{\mu\nu}. \label{emlag}
\end{equation}

The corresponding EMT reads

\begin{equation}
T_{\rm em \nu}^\mu = F_{\nu \alpha} F^{\mu \alpha} - \frac{1}{4}
F_{\alpha \beta} F^{\alpha\beta}. \label{emEMT}
\end{equation}

Since we consider the static spherically-symmetric configuration,
all the field functions should depend on the spatial variable $u$
only. Herewith
\begin{equation}
F_{10} (u) = - F_{01} (u) = A^\prime, \label{emc}
\end{equation}
with all other components $F_{\mu\nu}  \equiv 0.$ Here we assume
that the vector potential has only one nontrivial component $A_\mu =
(A,\,0,\,0,\,0).$

The electromagnetic field equation
\begin{equation}
\frac{1}{\sqrt{-g}} \frac{\partial}{\partial x^\nu}\left(\sqrt{-g}
F^{\nu\mu}\right) = 0, \label{emeq}
\end{equation}
in this case takes the form
\begin{equation}
\frac{\partial}{\partial u}\left(e^{2 \alpha} F^{10}\right) = 0,
\label{emeq1}
\end{equation}
with the solution
\begin{equation}
F^{10} = q e^{-2 \alpha} = q e^{-2 (\gamma + 2 \beta)}, \quad q =
{\rm const.} \label{sol}
\end{equation}
The foregoing equations leads to
\begin{equation}
A^\prime = - q e^{2\gamma}, \label{A1}
\end{equation}

The components of the EMT in this case takes the form
\begin{equation}
T_0^0 = T_1^1 = - T_2^2 = - T_3^3 = - \frac{1}{2} q^2 e^{-4 \beta}.
\label{emtrmc}
\end{equation}
In this case the Einstein system reads

\begin{subequations}
\label{CFhcem}
\begin{align}
\beta^\prime &= \nu, \label{CFem1}\\
\gamma^\prime &= \tau, \label{CFem2}\\
\nu^\prime &=  \nu^2 + 2 \nu \tau, \label{CFem3hc}\\
\tau^\prime &= - \kappa q^2 e^{-4\beta} . \label{CFem4hc}
\end{align}
\end{subequations}

The system \eqref{CFhcem} we solve numerically. $\kappa$ is taken to
be unity, whereas the initial values are set to be $\tau (0) =
0.3,\,\nu (0) = 0.2,\, \gamma (0) = 0.3,\, \beta (0) = 0.3,\, A (0)
= 1$. In Fig. \ref{metricemf}  we have drawn the picture of the
metric functions when the source is given by the electromagnetic
field. In Fig. \ref{vecpot} the corresponding vector potential is
demonstrated. In this case $\gamma (u)$ and $\beta (u)$ behave like
the one with radiation illustrated in Fig. \ref{metricpfrad}.

\vskip 1 cm \myfigures{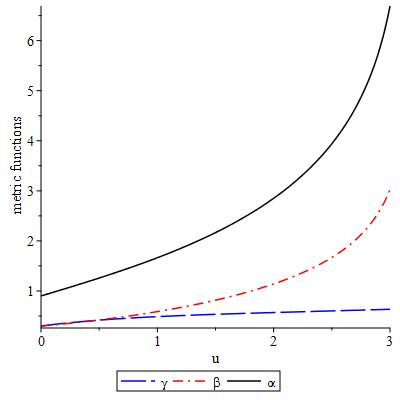}{0.46}{Plot of metric functions for
electromagnetic field}{0.45}{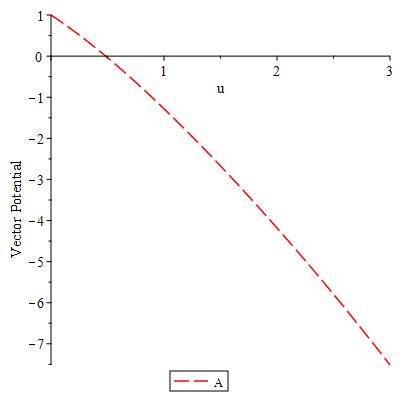}{0.43}{Plot of vector potential
$A(u)$}{0.45}

\vskip 5mm

{\bf Interacting scalar and electromagnetic fields}

\vskip 3mm

Let us finally consider the interacting scalar and spinor field
given by the Lagrangian \cite{Saha1997IJTP}

\begin{equation}
L = \frac{1}{2} \varphi_{,\alpha} \varphi^{,\alpha} - \frac{1}{4}
F_{\alpha\beta} F^{\alpha\beta}\Psi(\varphi), \quad \Psi(\varphi) =
1 + \lambda \Phi(\varphi). \label{lagint}
\end{equation}

Here $\lambda$ is the coupling constant between the scalar and
electromagnetic fields. Setting $\lambda = 0$ one obtains the case
with minimal coupling.

The corresponding equations
\begin{equation}
\partial_\nu \left(\sqrt{-g} g^{\mu\nu} \varphi_{,\mu}\right) +
\frac{1}{2} \sqrt{-g} F_{\alpha\beta} F^{\alpha\beta} \Psi_\varphi =
0, \label{scint}
\end{equation}
and
\begin{equation}
\partial_\nu \left(\sqrt{-g} F^{\mu\nu} \Psi(\varphi)\right) = 0,
\label{emint}
\end{equation}
where $\Psi_\varphi = d\Psi/d\varphi.$

The EMT in this case is

\begin{equation}
T_\nu^\mu = \left[\varphi_{,\nu} \varphi^{,\mu} - F_{\mu\beta}
F^{\nu\beta}\Psi(\varphi)\right]- \frac{1}{4} \delta_\nu^\mu \left[2
\varphi_{,\alpha} \varphi^{,\alpha} - F_{\alpha\beta}
F^{\alpha\beta}\Psi(\varphi)\right]. \label{emtint}
\end{equation}
The corresponding field equations read

\begin{equation}
\left(\sqrt{-{\tilde g}} g^{11} \varphi^\prime\right)^\prime +
\frac{1}{2} \sqrt{-{\tilde g}} F_{10} F^{10} \Psi_\varphi = 0,
\label{scint1}
\end{equation}
and
\begin{equation}
\left( \sqrt{-{\tilde g}}F^{10} \Psi(\varphi)\right)^\prime = 0,
\label{emint1}
\end{equation}
where $\sqrt{-{\tilde g}} =  \sqrt{- g}/\sin {\vartheta}.$

The solution to the \eqref{emint1} can be written as
\begin{equation}
F^{10} = q P(\varphi)/\sqrt{-{\tilde g}}, \quad  P(\varphi)= 1/
\Psi(\varphi), \quad q = {\rm const.} \label{emint2}
\end{equation}
From \eqref{emint2} one finds

\begin{equation}
A^\prime = - q e^{2\gamma} P(\varphi), \label{A2}
\end{equation}

On account of \eqref{emint2} the scalar field equation
\eqref{scint1} now looks
\begin{equation}
\varphi^{\prime\prime} = \frac{q^2}{2} e^{2\gamma} P_{\varphi}.
\label{scint2}
\end{equation}
The components of EMT in this case read
\begin{equation}
T_\nu^\mu = \frac{1}{2} e^{-2\alpha} \varphi^{\prime 2} {\rm diag}
\left(+1,\,-1,\,+1,\,+1\right) + \frac{q^2}{2} e^{2 (\gamma -
\alpha)} P^2 {\rm diag} \left(+1,\,+1,\,-1,\,-1\right).
\label{empcom}
\end{equation}

So in this case we have the following system of equations

\begin{subequations}
\label{CFhcint}
\begin{align}
\beta^\prime &= \nu, \label{CFint1}\\
\gamma^\prime &= \tau, \label{CFint2}\\
\nu^\prime &=  \nu^2 + 2 \nu \tau + \frac{\kappa}{2} e^{-2 (\gamma +
2 \beta)} \frac{1}{P^2}, \label{CFint3hc}\\
\tau^\prime &=  \frac{\kappa}{2} q^2 e^{-4 \beta} P^2 \label{CFint4hc}\\
\varphi^\prime &= \psi, \label{psiint}\\
\psi^\prime &= \frac{q^2}{2} e^{\gamma} P_\varphi. \label{scint}
\end{align}
\end{subequations}

As in previous cases we solve the system \eqref{CFhcint}
numerically. Again we set $\kappa = 1$ and $\tau (0) = 0.5,\, \nu
(0) = 0.5,\, \gamma (0) = 0.2,\, \beta (0) = 0.3,\, \varphi (0) =
1,\, \varphi^\prime (0) = 0.2,\, A (0) = 0.2$. In Figs.
\ref{metricint} and \ref{gammaint} we we have plotted the metric
functions and $\gamma (u)$, respectively for $P = J^{2-4/\sigma}
\left(1 - J^{2/\sigma}\right)$ with $J = \lambda \varphi$, $\sigma =
2 n + 1$ and $n = 1,\,2,\,3....$. Here we set $n = 1$ and $\lambda =
1$. Comparing the Figs. \ref{metricscalar}, \ref{metricemf} and
\ref{metricint} one sees, in this case the the behavior of the
metric functions determined by the electromagnetic field.

\vskip 1 cm \myfigures{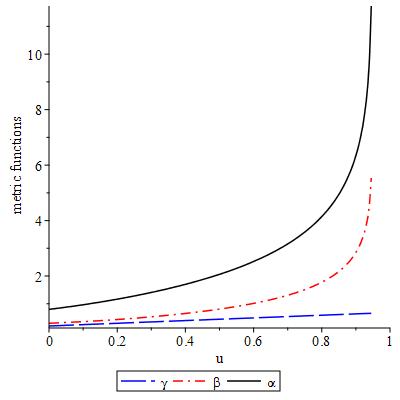}{0.46}{Plot of metric functions for
electromagnetic field}{0.45}{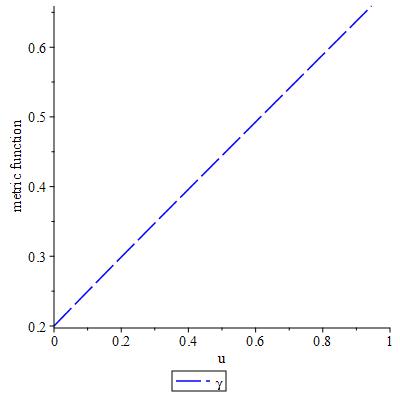}{0.43}{Plot of metric function
$\gamma(u)$}{0.45}

In Figs. \ref{scalarint} and \ref{vectorint} corresponding picture
of the scalar and electromagnetic fields are illustrated.

\vskip 1 cm \myfigures{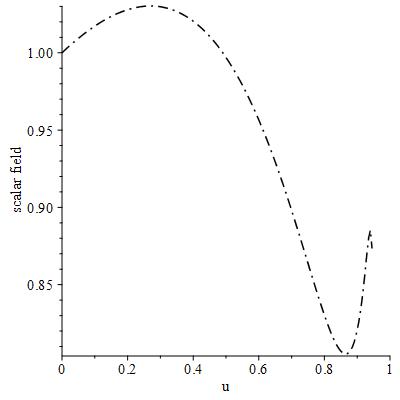}{0.46}{Plot of scalar functions for
interacting scalar and electromagnetic
fields}{0.45}{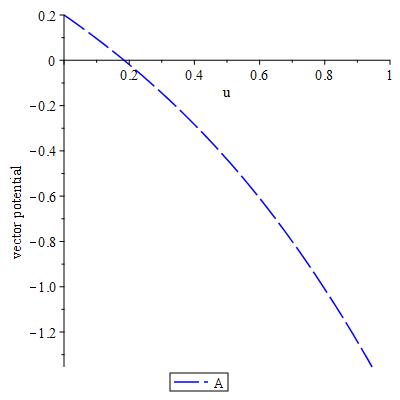}{0.43}{Plot of vector potential for
interacting scalar and electromagnetic fields}{0.45}

\section{Conclusion}

We have considered a static spherically symmetric space-time and
analyzed it in presence of different types of matter. The system was
transformed in such a way that some idea about the qualitative
solutions to the Einstein field equations can be made looking at the
type of source field, precisely the interrelation  between the
components of EMT. For simplicity we have considered only harmonic
coordinate. We plan to extend our study for the other cases as well
as for some other realistic source fields in near future.

\vskip 7mm

\noindent {\bf Acknowledgments}\\
The publication was prepared with the support of the "RUDN
University Program 5-100" and also partly supported by a joint
Romanian-JINR, Dubna Research Project, Order no.396/27.05.2019 p-71.


\begin{thebibliography}{9999}

\bibitem{BronBook} {\it Bronnikov K.A. and Rubin S.G.} Black Holes,
Cosmology and extra Dimensions. (World Scientific, Singapore, 2013)

\bibitem{Saha2018} {\it Saha B.} Eur. Phys. J. Plus {\bf 133} 461 (2018)

\bibitem{BRS2020} {\it Bronnikov K.A., Rybakov Yu.P., and Saha B.}
Eur.Phys. J. Plus {\bf 135} 124 (2020)

\bibitem{Bron2020} {\it Bronnikov K.A. et. al.} Rastall's theory of gravity: Spherically symmetric solutions and
the stability problem. Arxiv: 2007.01945V1 [gr-qc] (2020)

\bibitem{shikin} {\it Shikin G.N.} Fundamentals of the theory of
solitons in general relativity. (URSS, Moscow, 1995)


\bibitem{Saha1997IJTP}  {\it Rybakov, Yu. P., Saha B. and Shikin, G.N.}
Int. J. Theor. Phys. {\bf 36}, 1475 (1997)


\end{thebibliography}
\end{document}